\documentclass[aps,prl,twocolumn,superscriptaddress]{revtex4}
\usepackage{epsf,graphicx}
\usepackage{amssymb}
\usepackage{amsmath}
\usepackage{latexsym,bm,array,amsfonts,multirow}
\usepackage{color}
\usepackage{ulem}
\begin{document}

\title{High-$T_c$ superconductivity by mobilizing local spin singlets and\\ possible route to higher $T_c$ in pressurized La$_3$Ni$_2$O$_7$}
\author{Qiong Qin}
\affiliation{Beijing National Laboratory for Condensed Matter Physics and Institute of
	Physics, Chinese Academy of Sciences, Beijing 100190, China}
\affiliation{University of Chinese Academy of Sciences, Beijing 100049, China}
\author{Yi-feng Yang}
\email[]{yifeng@iphy.ac.cn}
\affiliation{Beijing National Laboratory for Condensed Matter Physics and Institute of
Physics, Chinese Academy of Sciences, Beijing 100190, China}
\affiliation{University of Chinese Academy of Sciences, Beijing 100049, China}
\affiliation{Songshan Lake Materials Laboratory, Dongguan, Guangdong 523808, China}
\date{\today}

\begin{abstract}
We clarify the pairing mechanism of high-$T_c$ superconductivity in bilayer La$_3$Ni$_2$O$_7$ under high pressure by employing the static auxiliary field Monte Carlo approach to simulate a minimal effective model that contains local $d_{z^2}$ interlayer spin singlets and metallic $d_{x^2-y^2}$ bands. Superconductivity is induced when the local spin singlet pairs are mobilized and attain long-distance phase coherence by hybridization with the metallic bands. When projected onto realistic Fermi surfaces, it yields a nodeless $s$-wave gap on the $\gamma$ Fermi surface, and extended $s$-wave gaps of the same (opposite) sign on the $\alpha$ ($\beta$) Fermi surface due to its bonding (antibonding) character, with nodes or gap minima along the diagonal direction of the two-dimensional Brillouin zone. We find a dual role of the hybridization that not only induces global phase coherence but also competes with the spin singlet formation. This lead to a tentative phase diagram where $T_c$ varies nonmonotonically with the hybridization, in good correspondence with experimental observations. A roughly linear relation is obtained for realistic hopping and hybridization parameters: $T_c\approx 0.04-0.05 J$, where $J$ is the interlayer superexchange interaction. We emphasize the peculiar tunability of the bilayer structure and propose that $T_c$ may be further enhanced by hole doping or applying uniaxial pressure along the $c$ axis on superconducting La$_3$Ni$_2$O$_7$. Our work provides reliable numerical evidence for the pairing mechanism of high-$T_c$ superconductivity in La$_3$Ni$_2$O$_7$ and points out a potential route to achieve even higher $T_c$.
\end{abstract}
\maketitle

\textit{Introduction.} The recently discovered high-$T_c$ superconductivity in the bilayer nickelate La$_3$Ni$_2$O$_7$ under high pressure \cite{Sun2023b,Liu2023a,Hou2023,Zhang2023c} has stimulated intensive interest concerning its basic electronic structures \cite{Luo2023,Zhang2023b,Christiansson2023,Shilenko2023,Wu2023a,Cao2023,Chen2023} and possible pairing mechanism \cite{Yang2023a,Lechermann2023,Sakakibara2023,Gu2023,Shen2023,Liu2023b,Lu2023c,Zhang2023d,Oh2023,Liao2023,Qu2023,Yang2023b,Jiang2023a,Huang2023,Zhang2023e}. While first-principles band calculations have predicted a Ni-$d^{7.5}$ configuration with an almost fully filled $d_{z^2}$ bonding band and two $d_{x^2-y^2}$ bands near quarter filling, it has also been argued that this weak-coupling picture is not enough to explain the high $T_c$ of about 80 K \cite{Yang2023b}. Indeed, strongly correlated electronic structure calculations have revealed well-formed $d_{z^2}$ moments with a large interlayer superexchange interaction $J$ via the O-$p_z$ orbital \cite{Cao2023}. This lays the basis for a strong-coupling picture, where the $d_{z^2}$ electrons provide preformed interlayer spin singlets with a large pairing energy and the metallic $d_{x^2-y^2}$ bands provide a large phase stiffness. While it was suggested that a strong coupling of the two components could give rise to high $T_c$ \cite{Berg2008,Yang2023b}, other weak-coupling scenarios have also been put forward to explain the pairing. It is therefore urgent to give more concrete calculations for qualitative or even quantitative comparisons with experimental observations. In addition, one may be curious if higher $T_c$ can be achieved in La$_3$Ni$_2$O$_7$ by proper tuning besides hydrostatic pressure.

In this Letter, we propose that the high-$T_c$ superconductivity arises by mobilizing the local spin singlets of $d_{z^2}$ electrons by hybridization with metallic $d_{x^2-y^2}$ bands \cite{Yang2023b} and provide detailed numerical support for this pairing mechanism by performing static auxiliary field Monte Carlo simulations \cite{Mayr2005a,Dubi2007,Pasrija2016,Karmakar2020,Dong2021a,Mukherjee2014,Liang2013} on a minimal effective low-energy model. We construct a phase diagram showing a qualitatively similar nonmonotonic evolution of $T_c$ with increasing hybridization strength as observed in experiments under pressure tuning. Our calculations reveal a dual role of the hybridization in driving the superconductivity. On the one hand, it helps to mobilize the local spin singlet pairs and induce a global  phase coherence for the superconductivity; on the other hand, it competes with the interlayer superexchange interaction and tends to suppress the pairing strength. The overall good consistency with the experiments provides a strong support of our scenario for the high-$T_c$ superconductivity in La$_3$Ni$_2$O$_7$ under high pressure. We further find a roughly linear relation for realistic hopping and hybridization parameters, $T_c\approx 0.04-0.05 J$, and propose that higher $T_c$ may be achieved by further applying uniaxial pressure along the $c$ axis on superconducting La$_3$Ni$_2$O$_7$. Our work points out that mobilizing preformed spin singlets may be a general route for pursuing more high-$T_c$ superconductors. 

\textit{Method.} We focus only on the pairing mechanism and study how superconductivity emerges based on the following minimal effective bilayer Hamiltonian \cite{Yang2023b},
\begin{equation}
\begin{split}
H&=J\sum_{i}\bm{S}_{1i}\cdot\bm{S}_{2i}  -\sum_{l\langle ij\rangle\sigma}V_{ij}\left(d_{li\sigma}^{\dagger}c_{lj\sigma}+h.c\right)\\
&-\sum_{l\langle ij\rangle \sigma}(t_{ij}+\mu\delta_{ij})c_{li\sigma}^{\dagger}c_{lj\sigma},
\end{split}
\end{equation} 
where $d_{li\sigma}$ ($c_{li\sigma}$) is the annihilation operator of the $d_{z^2}$ ($d_{x^2-y^2}$) electrons  with spin $\sigma$ on site $i$ of layer $l$, and $\bm{S}_{li}=\frac12\sum_{ss'}d_{lis}^{\dagger}\bm{\sigma}_{ss'}d_{lis'}$ is the spin density operator of $d_{z^2}$ electrons. The minimal model only includes the inter-layer antiferromagnetic superexchange interaction $J$ for $d_{z^2}$ electrons, the nearest-neighbor hopping parameter $t_{ij}$ and the chemical potential $\mu$ of the itinerant $d_{x^2-y^2}$ electrons, and the in-plane nearest-neighbor hybridization $V_{ij}$ between two orbitals. Because the $d_{z^2}$ and $d_{x^2-y^2}$ wave functions are orthogonal on the same Ni ion, the hybridization occurs mainly between nearest-neighbor sites and has opposite signs along the $x$ and $y$ directions ($V_{i,i+x}=-V_{i,i+y}=V$). The interlayer hopping is also dropped, which may affect the Fermi surfaces and their gap structures but does not change the basic pairing mechanism. All other parameters are either small or strongly renormalized by electronic correlations \cite{Luo2023,Cao2023,Yang2023b}, and thus play no significant role in the superconductivity. In particular, the intralayer superexchange interactions are negligible for both orbitals due to the quarter filling of $d_{x^2-y^2}$ as in heavily hole-doped cuprates and the weak intralayer hybridization of $d_{z^2}$ with the O-$p_{x/y}$ orbitals.

\begin{figure}[t]
	\begin{center}
		\includegraphics[width=8cm]{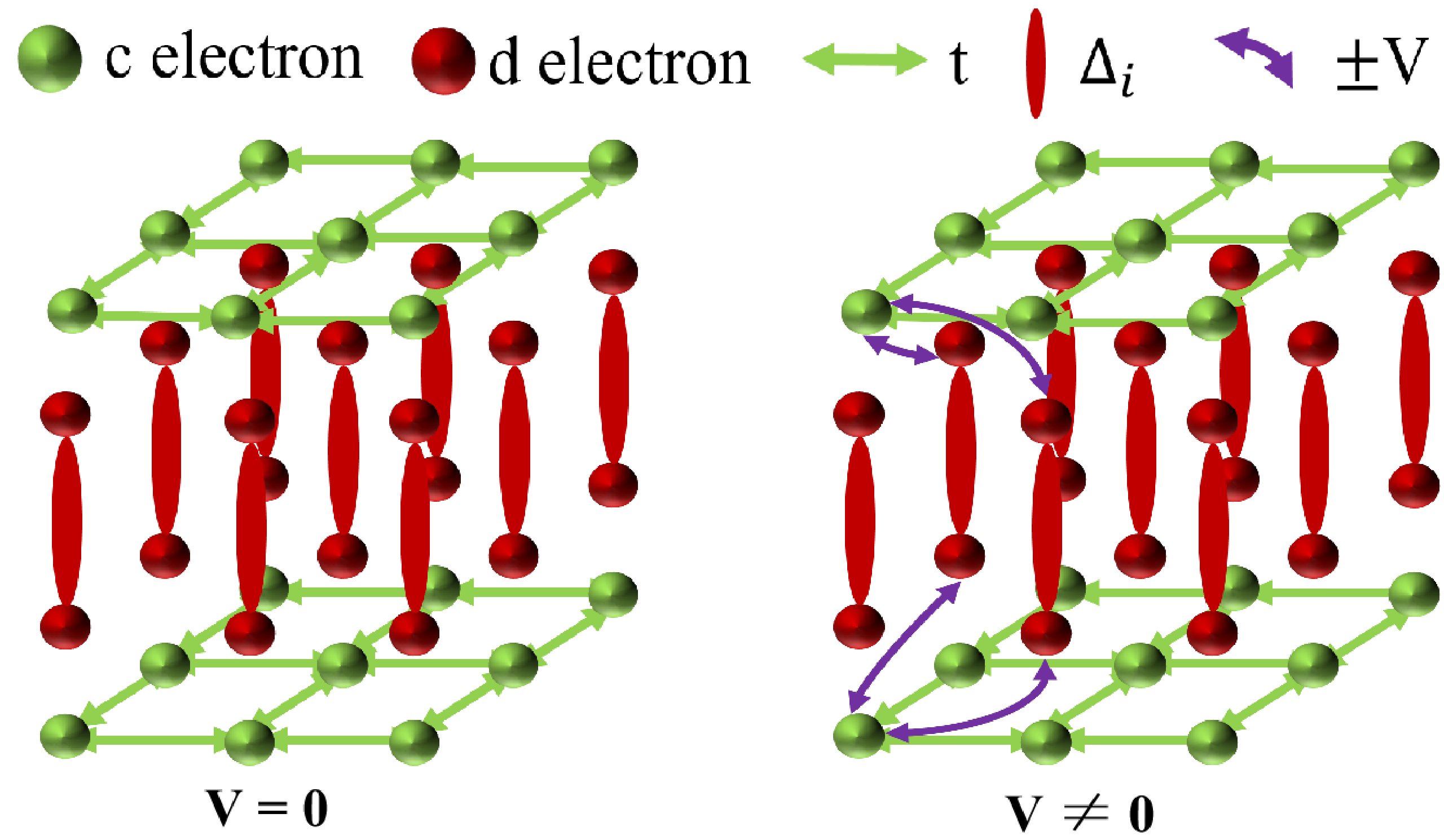}
	\end{center}
	\caption{Illustration of local $d_{z^2}$ interlayer spin singlets ($d$ electron) getting mobilized and attaining phase coherence by nearest-neighbor hybridization with metallic $d_{x^2-y^2}$ bands ($c$ electron). For clarity, the $c$ orbitals are all shifted outwards.}
	\label{fig1}
\end{figure}

Figure \ref{fig1} gives an illustration of the above minimal effective model. Although simple, this model covers all essential ingredients for the superconductivity and gives a minimal description of $d_{z^2}$ spin singlet pairs mediated by the interlayer antiferromagnetic superexchange coupling. The absence of direct hopping between $d_{z^2}$ orbitals indicates that their spin singlets are local and cannot by themselves attain the phase coherence to reach superconductivity. Superconductivity can only emerge and become established by hybridization with the metallic $d_{x^2-y^2}$ bands. To see how this mechanism is realized, we first decouple the superexchange term via the Hubbard-Stratonovich transformation \cite{Coleman2015}, 
\begin{equation}
J\bm{S}_{1i}\cdot\bm{S}_{2i}\rightarrow \sqrt{2}\bar{\Delta}_i\psi_{i}+\sqrt{2}\bar{\psi}_{i}\Delta_i+\frac{8\bar{\Delta}_i\Delta_i}{3J},
\end{equation}
where $\psi_{i}=\frac{1}{\sqrt{2}}\left(d_{1i\downarrow}d_{2i\uparrow}-d_{1i\uparrow}d_{2i\downarrow}\right)$ represents the local interlayer spin singlet of $d_{z^2}$ electrons at site $i$ and $\Delta_i$ is the corresponding auxiliary pairing field. However, direct Monte Carlo simulations typically suffer from a severe negative sign problem. To avoid this, we ignore the temporal dependence and adopt a static approximation, $\Delta_i(\tau)\rightarrow \Delta_i$. The fermionic degrees of freedom have a bilinear form and can be easily integrated out. Following Ref. \cite{Qin2023PRB}, this gives an effective action $S_{\rm eff}(\Delta_i)$ that depends solely on the complex pairing fields and can be simulated without the negative sign problem.

\begin{figure}[t]
	\begin{center}
		\includegraphics[width=8.6cm]{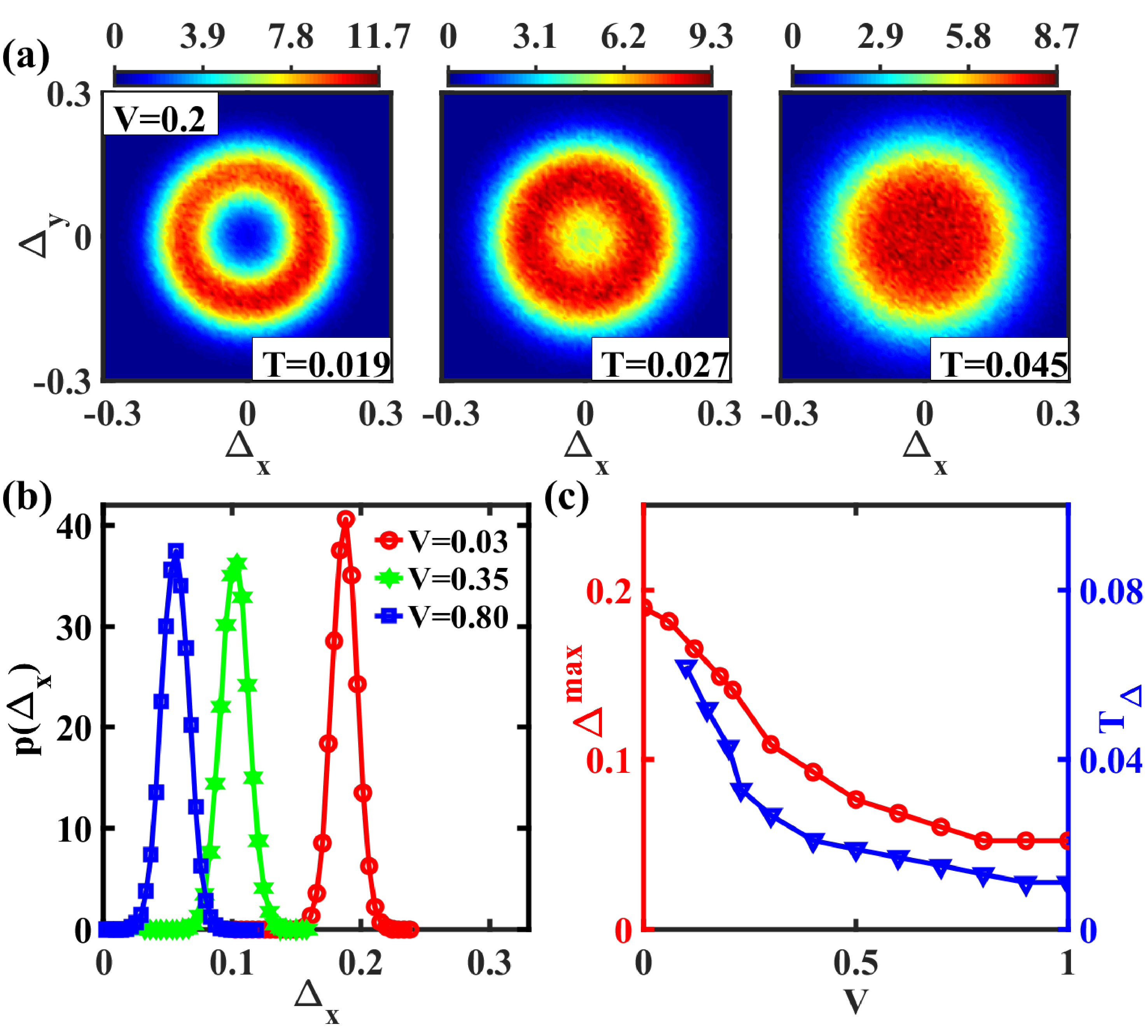}
	\end{center}
	\caption{(a) Intensity plot of the probabilistic distribution $p(\Delta)$ of the local pairing fields on the complex plane $\Delta=(\Delta_x,\Delta_y)$ for different temperatures $T=0.019$, 0.027, 0.045 at a fixed hybridization $V=0.2$. (b) Probabilistic distribution $p(\Delta_x)$ within a narrow cut $|\Delta_y|\le0.01$ for $V=0.03$, 0.35, 0.8 at a low temperature $T=0.001$. (c) Evolution of the peak position $\Delta^{\rm max}$ of $p(\Delta_x)$ at the low-temperature limit and the onset temperature $T_{\Delta}$ of local spin singlets as functions of the hybridization $V$. $T_\Delta$ marks the transition from the ring to a single maximum at the origin in the distribution plot (a). $t$ is set to unity as the energy unit and $J=0.5$.}
	\label{fig2}
\end{figure}

The static auxiliary field Monte Carlo method has been verified in previous studies of unconventional superconductivity \cite{Qin2023PRB,Han2010,Zhong2011,Singh2021,Mayr2005a,Dubi2007,Karmakar2020}. It ignores dynamical fluctuations of the pairing fields but captures well their thermal and spatial fluctuations. The static approximation breaks down at extremely low temperatures or near quantum phase transitions, but is suitable in our case to study how the phase coherence is established at finite temperature for the local $d_{z^2}$ spin singlets \cite{Dong2022PRB,Qin2023PRB}. We perform the Monte Carlo simulations on a 10$\times$10 bilayer lattice with periodic boundary conditions. Our results are examined on larger lattices and remain robust due to the local nature of the $d_{z^2}$ interlayer pairing. Hereafter, we set $t=1$ as the energy unit, and choose $J=0.5t$ for the superexchange coupling as estimated from the tight-binding parameters \cite{Wu2023a,Yang2023b,Luo2023}. For simplicity, the chemical potential is fixed to $\mu=-1.3$ so that the $d_{x^2-y^2}$ and $d_{z^2}$ orbitals are near quarter and half filled, respectively. The effective hybridization strength $V$ may vary with pressure \cite{Yang2023b} and is therefore taken as a free parameter to construct the superconducting phase diagram.

\textit{Local spin singlet pairs.} We first study the probabilistic distribution of the local spin singlets, $p(\Delta_i)=Z^{-1}e^{-S_{\rm eff}(\Delta_i)}$, where $Z$ is the partition function playing the role of the normalization factor. A typical result is plotted in Fig. \ref{fig2}(a) on the complex plane $(\Delta_x,\Delta_y)$ for three different temperatures at $V=0.2$. We find the distribution clusters around the origin at high temperatures but gradually develops into a ring below a characteristic temperature $T_\Delta$. The finite radius of the ring marks the formation of local $d_{z^2}$ spin singlet pairs. At low temperatures, its value reflects the intrinsic pairing strength and may be estimated by plotting the distribution $p(\Delta_x)$ within a narrow cut $|\Delta_y|\le0.01$. This is plotted in Fig. \ref{fig2}(b) for $T=0.001$ and determined by the maximum of the distribution. Interestingly, the peak position moves to a smaller $\Delta_x$ with increasing hybridization $V$, implying a reduced pairing strength for strong hybridization. This is best seen in Fig. \ref{fig2}(c), where we plot $T_\Delta$ and $\Delta^{\rm max}$ as functions of $V$. Both quantities decrease monotonically and reveal the competition between the local spin singlet formation and the hybridization. 

\begin{figure}[ptb]
	\begin{center}
		\includegraphics[width=8.6cm]{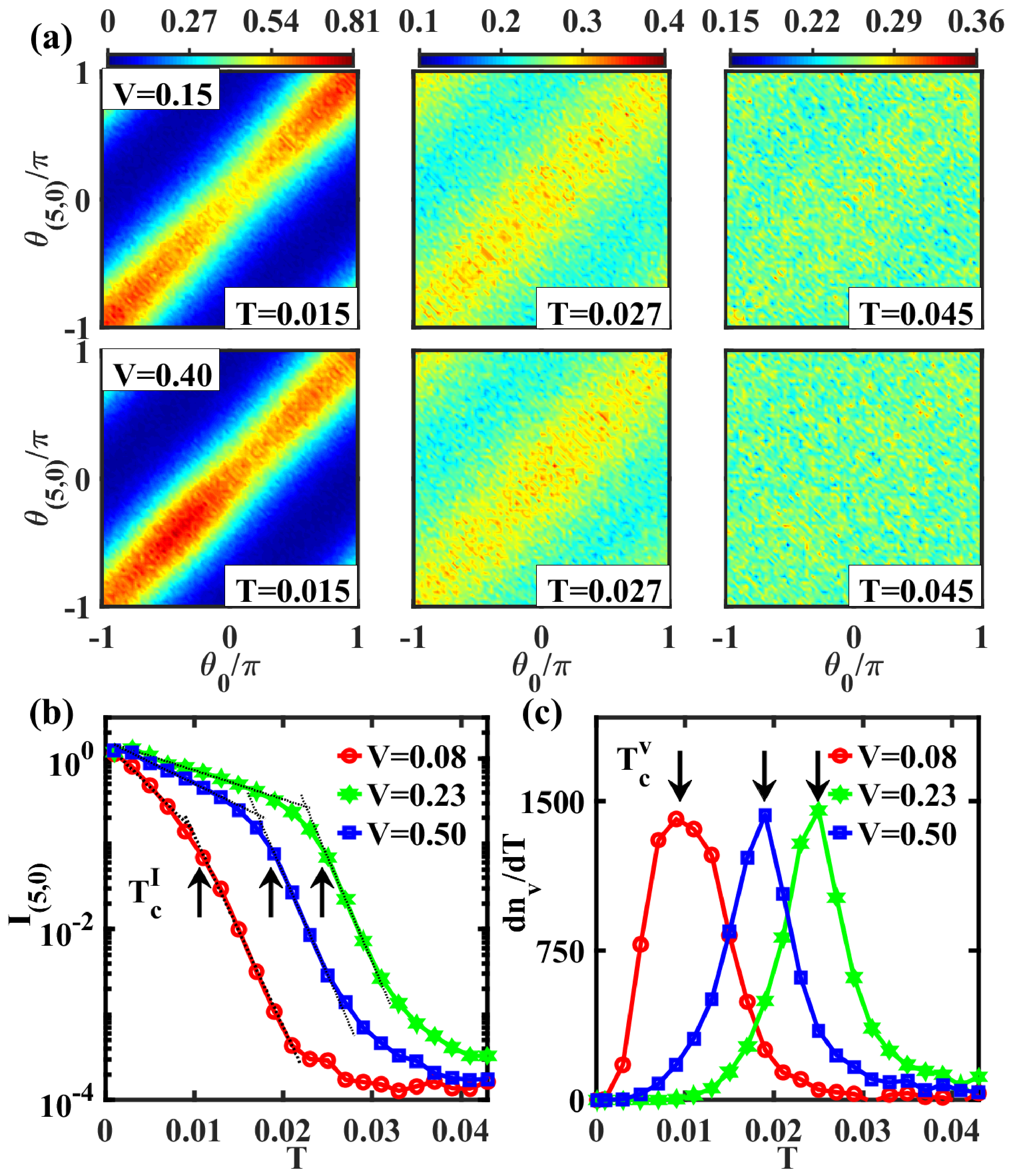}
	\end{center}
	\caption{(a) Intensity plot of the joint distribution between $\theta_{\bm{0}}$ and $\theta_{(5,0)}$ for $V=0.15$ (upper panels) and 0.4 (lower panels) at three different temperatures, showing a similar evolution from a uniform distribution to a stripe pattern. (b) The corresponding phase mutual information $I_{(5,0)}$ as a function of temperature for different values of $V$. The slope change at low temperatures marks the crossover from short- to long-distance phase correlations and defines the temperature scale $T_c^{\rm I}$. (c) Temperature dependence of  $dn_{\rm v}/dT$ for different hybridization strengths. The maximum reflects the characteristic BKT transition for two-dimensional superconductivity and defines the temperature scale $T_c^{\rm v}$.}
	\label{fig3}
\end{figure}

\textit{Phase coherence.} To see how superconductivity emerges from these local spin singlets, we study the long-distance phase correlations of the complex pairing fields $\Delta_i\equiv |\Delta_i|e^{i\theta_i}$. Figure \ref{fig3}(a) shows the joint distribution of the phase $\theta_i$ on two distant bonds, where $\theta_{\bm{0}}$ is located on a chosen origin and $\theta_{(5,0)}$ is on the bond at a distance $\mathbf{R}=(5,0)$. With lowering temperature, we see the evolution from a uniform distribution at $T=0.045$ to a stripe feature at $T=0.015$. This indicates the gradual development of phase correlations between two distant bonds, a signature of global phase coherence between local spin singlets. For comparison, we show the results for $V=0.15$ and 0.4. They have very different $T_\Delta=0.052$ and 0.021, but the patterns of the joint phase distributions look quite similar for the same temperature. There is an obvious disparity between the spin singlet formation and long-distance phase correlations. 

To clarify this, we quantify the phase correlations by introducing the phase mutual information \cite{Cover2006,Kraskov2004PRE,Varanasi1999,Khan2007,Belghazi2018PMLR,Poolel2019PMLR},
\begin{eqnarray}
	I_{\bm{R}}=\int d\theta_{\bm{0}} d\theta_{\bm{R}}~ p(\theta_{\bm{0}},\theta_{\bm{R}})\ln\frac{p(\theta_{\bm{0}},\theta_{\bm{R}})}{p(\theta_{\bm{0}})p(\theta_{\bm{R}})},
	\end{eqnarray}
where $p(\theta_i)$ is the marginal distribution of the phase $\theta_i$ at site $i$ and $p(\theta_{\bm{0}},\theta_{\bm{R}})$ is their joint probabilistic distribution on two distant bonds at $\bm{0}$ and $\bm{R}$ after integrating out the pairing amplitude $|\Delta_i|$. Figure \ref{fig3}(b) plots the phase mutual information $I_{(5,0)}$ as functions of the temperature for three different values of $V$. In all cases, we find a gradual increase of the phase mutual information with lowering temperature. The increase grows rapidly in an intermediate-temperature range, marking a rapid development of phase correlations on two distant bonds. At a lower temperature $T_c^{\rm I}$, a slope change is seen below which the phase mutual information grows less rapidly and seems to saturate towards some zero-temperature limit. We will see that $T_c^{\rm I}$ may be identified as the superconducting transition temperature, at which the phase coherence is established between local spin singlet pairs on distant bonds. 

To further establish the superconducting transition, we also calculate the vortex number \cite{Drouin-Touchette2022}, $n_{\rm v}=\sum_{i}\langle \delta_{w_i,1}\rangle$, where the average is for all pairing configurations and $w_i$ is the winding number for $\theta_i\rightarrow\theta_{i+\hat{x}}\rightarrow\theta_{i+\hat{x}+\hat{y}}\rightarrow\theta_{i+\hat{y}}\rightarrow\theta_i$. We find $n_{\rm v}$ increases rapidly in an intermediate-temperature range. Its derivative $dn_{\rm v}/dT$ is shown in Fig. \ref{fig3}(c) and defines another temperature scale $T_c^{\rm v}$ at the maximum. Following the picture of the Berezinskii-Kosterlitz-Thouless (BKT) transition for two-dimensional superconductivity \cite{Berezinskii1972,Kosterlitz1973,Kosterlitz1974}, the vortex-antivortex pairs are excited with temperature and become unbound across the transition, causing a rapid increase of $n_{\rm v}$ around $T_c$. The peak in $dn_{\rm v}/dT$ therefore marks a characteristic feature of the BKT transition. Intriguingly, with increasing $V$, the peak position moves first towards higher temperatures ($V=0.23$) but then backwards to lower temperatures ($V=0.5$), indicating a nonmonotonic variation of $T_c^{\rm v}$ in contrast to $T_\Delta$.

\begin{figure}[t]
	\begin{center}
		\includegraphics[width=8cm]{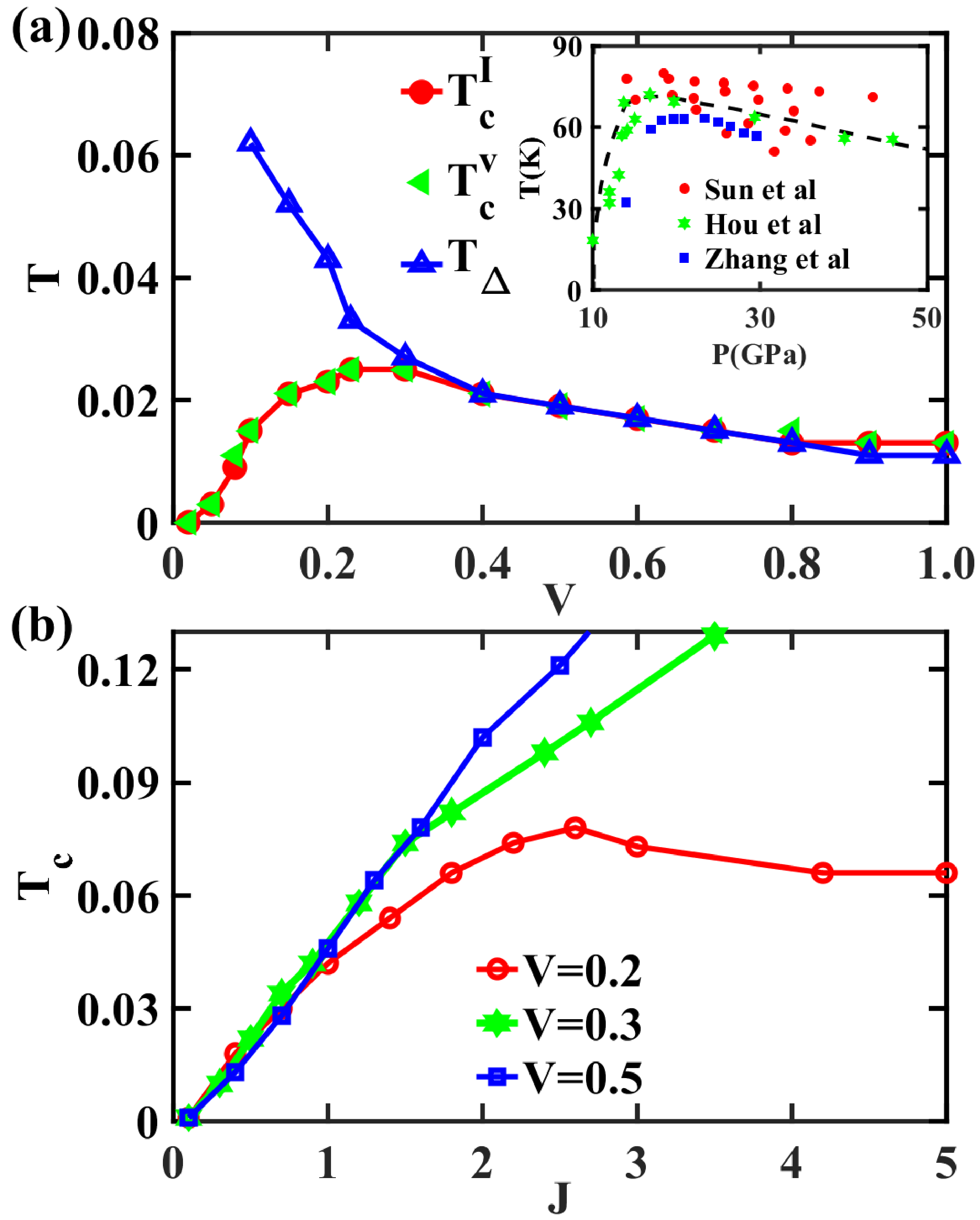}
	\end{center}
	\caption{ (a) Theoretical phase diagram of the superconductivity, showing all three temperature scales: $T_c^{\rm I}$ from the slope change marking the long-distance phase coherence in the phase mutual information plotted in Fig. \ref{fig3}(b), $T_c^{\rm v}$ from the maximum of $dn_{\rm v}/dT$ marking the BKT transition in Fig. \ref{fig3}(c), and $T_{\Delta}$ from the probabilistic distribution of local pairing fields marking the transition from the ring distribution to a single maximum at the origin in Fig. \ref{fig2}(c). All results are obtained for $J=0.5$. For comparison, the inset reproduces the measured $T_c$ in experiments under pressure on different samples \cite{Sun2023b,Hou2023,Zhang2023c}, where the dashed line is a guide to the eye. (b) Variation of $T_c$ estimated from $T_c^{\rm v}$ as a function of the superexchange interaction $J$ for $V=0.2$, 0.3, 0.5. $t$ is taken as the energy unit.}
	\label{fig4}
\end{figure}

\textit{Superconducting phase diagram.} For better comparison, we construct a superconducting phase diagram on the $V$-$T$ plane in Fig. \ref{fig4}(a) and plot all three temperature scales, $T_\Delta$, $T_c^{\rm I}$, $T_c^{\rm v}$ as functions of the hybridization parameter $V$. Indeed, while $T_\Delta$ decreases continuously with increasing $V$, both $T_c^{\rm I}$ and $T_c^{\rm n}$ vary nonmonotonically and collapse roughly on the same curve. The excellent coincidence between $T_c^{\rm I}$ and $T_c^{\rm v}$ provides further support for the superconducting transition through global phase coherence of local spin singlets and gives a consistent definition of $T_c$. We find a maximum $T_c\approx 0.025$ at $V\approx 0.25$. For smaller hybridization, $T_c$ and $T_\Delta$ behave oppositely and there exists a wide intermediate temperature region $T_c<T<T_\Delta$ where local spin singlet pairs exist but show no long-distance phase coherence. This marks a region of preformed pairs as previously proposed for underdoped cuprates \cite{Emery1995,Keimer2015}. We obtain the ratio $2\Delta^{\rm max}/T_{\Delta}\approx 4-6$, a value close to those of pseudogaps observed in many experiments \cite{Yoshida2009,Richter2013}. Superconductivity is only established when the local $d_{z^2}$ spin singlet pairs are get mobilized and attain phase coherence through hybridization with metallic $d_{x^2-y^2}$ bands.

For larger hybridization, the three temperature scales behave quantitatively similar, indicating that the superconductivity is now constrained by the spin singlet formation rather than the phase coherence. The decrease of $T_c$ with increasing $V$ reflects the suppression of the pairing strength by the hybridization. The obtained ratio $2\Delta^{\rm max}/T_c \approx 7.5-9$ is commonly observed in many unconventional superconductors \cite{Kim2022,Yao2019,Inosov2011}, and may be examined in future experiments for superconducting La$_3$Ni$_2$O$_7$. 

It should be noted that for two-dimensional superconductivity, there is always a finite precursor region above $T_c$. This is not plotted in our phase diagram but corresponds to the narrow region above $T_c^{\rm I}$ in Fig. \ref{fig3}(b) where the phase mutual information grows rapidly with lowering temperature. In this regard, $T_\Delta$ estimated from Fig. \ref{fig2}(a) somewhat underestimates the onset temperature of the spin singlet pairs because of the large broadening of the ring. We will not go into more details on this because we are mainly interested in the behavior of $T_c$ and its comparison with experiments. 

Overall, our derived $V$-$T$ phase diagram resembles closely those observed in experiments for La$_3$Ni$_2$O$_7$ under pressure \cite{Sun2023b,Hou2023,Zhang2023c}, where $T_c$  exhibits a nonmonotonic variation: It increases rapidly to near 80 K from 10 to 18 GPa and then decreases gradually to about 50 K at 50 GPa as shown in the inset of Fig. \ref{fig4}(a). Our calculations suggest that this arises from a dual role of the hybridization, which mobilizes the local spin singlet pairs to induce global phase coherence but at the same time competes to suppress their pairing strength. It may also be illuminating to make some quantitative estimate for direct comparisons. Taking $t\approx0.5$ eV from first-principles calculations \cite{Luo2023}, our phase diagram for $J/t=0.5$ yields a maximum $T_c\approx 0.025t$, which is roughly 140 K, the same order of magnitude as the experimental $T_c$ near 80 K, considering that the real $T_c$ may be reduced by other factors beyond our minimal effective model. This overall agreement provides strong support for our theory.

\textit{Pairing symmetry.} Starting from the primary $d_{z^2}$ local interlayer spin-singlet pairing, superconducting gap structures can be directly derived using the tight-binding Hamiltonian for any realistic Fermi surfaces depending on their respective orbital and bonding characters. Electronic band structure calculations have yielded a small hole pocket ($\gamma$) from the $d_{z^2}$ bonding band, an electronlike Fermi surface ($\alpha$) from the hybridized $d_{x^2-y^2}$ and $d_{z^2}$ bonding bands, and a holelike Fermi surface ($\beta$) from the hybridized $d_{x^2-y^2}$ and $d_{z^2}$ antibonding bands  \cite{Luo2023,Cao2023}. Our theory then predicts a nodeless $s$-wave gap on $\gamma$, and extended $s$-wave gaps of the same sign on $\alpha$ and opposite sign on $\beta$. The $\alpha$ and $\beta$ gaps have nodes (or gap minima) along the diagonal direction of the two-dimensional Brillouin zone but grow rapidly away from the zone diagonal due to the anisotropic $d_{x^2-y^2}-d_{z^2}$ hybridization in momentum space, $V_\mathbf{k}\propto (\cos\mathbf{k}_x-\cos\mathbf{k}_y)$.

\textit{Route to higher $T_c$.} Given the observed maximum $T_c$ near 80 K in experiments, it is desirable to ask if higher $T_c$ may be achieved upon proper tuning. Despite some delicacy in the pressure or hybridization tuning, some insight may still be gained by taking liberties with the model calculations. To explore other possibilities, we fix the hybridization and change the superexchange interaction $J$. As shown in Fig. \ref{fig4}(b) for $V=0.2$, $T_c$ exhibits similar nonmonotonic behavior with increasing $J$. Evidently, the increase of $T_c$ at small $J$ is owing to the increase of the pairing strength, while its decrease at large $J$ is constrained by the phase coherence due to hybridization. The maximum $T_c$ can indeed be enhanced by tuning $J$. For small $J$, Fig. \ref{fig4}(b) suggests a roughly linear relation, $T_c\approx 0.04-0.05J$, for realistic values of $V$ and $t$. A smaller prefactor may be possible if $V$ is too small. A crude estimate for superconducting La$_3$Ni$_2$O$_7$ yields $J\approx 0.5$, which falls exactly in this region. Thus, a higher $T_c$ may be achieved by simply increasing the superexchange interaction $J$ at fixed $t$ and $V$. Fascinatingly, this might actually be realized in experiment by further applying uniaxial pressure along the $c$ axis, since the hopping $t$ and the hybridization $V$ are both in-plane parameters while $J$ is the interlayer coupling. The fact that they may be tuned separately highlights the importance of the bilayer structure of superconducting La$_3$Ni$_2$O$_7$. In addition, hole doping may also promote the effective $V$ and enhance $T_c$, possibly even at ambient pressure. We suggest future experiments to verify these simple but important predictions of our minimal effective model.

We thank Guang-Ming Zhang and Fu-Chun Zhang for stimulating discussions. This work was supported by the National Natural Science Foundation of China (Grants No. 11974397 and No. 12174429), the Strategic Priority Research Program of the Chinese Academy of Sciences (Grant No. XDB33010100), and the National Key Research and Development Program of China (Grant No. 2022YFA1402203).


\begin{thebibliography}{99}
	
	\bibitem{Sun2023b}H. Sun, M. Huo, X. Hu, J. Li, Z. Liu, Y. Han, L. Tang, Z. Mao, P. Yang, B. Wang, J. Cheng, D.-X. Yao, G.-M. Zhang, and M. Wang, Signatures of superconductivity near 80 K in a nickelate under high pressure, Nature \textbf{621}, 493 (2023).
	\bibitem{Liu2023a}Z. Liu, M. Huo, J. Li, Q. Li, Y. Liu, Y. Dai, X. Zhou, J. Hao, Y. Lu, M. Wang, and H.-H. Wen, Electronic correlations and energy gap in the bilayer nickelate La$_{3}$Ni$_{2}$O$_{7}$, arXiv:2307.02950.
	\bibitem{Hou2023}J. Hou, P. T. Yang, Z. Y. Liu, J. Y. Li, P. F. Shan, L. Ma, G. Wang, N. N. Wang, H. Z. Guo, J. P. Sun, Y. Uwatoko, M. Wang, G.-M. Zhang, B. S. Wang, and J.-G. Cheng, Emergence of high-temperature superconducting phase in the pressurized La$_3$Ni$_2$O$_7$ crystals, arXiv:2307.09865.
	\bibitem{Zhang2023c}Y. Zhang, D. Su, Y. Huang, H. Sun, M. Huo, Z. Shan, K. Ye, Z. Yang, R. Li, M. Smidman, M. Wang, L. Jiao, and H. Yuan, High-temperature superconductivity with zero-resistance and strange metal behavior in La$_{3}$Ni$_{2}$O$_{7}$, arXiv:2307.14819.
	
	\bibitem{Luo2023}Z. Luo, X. Hu, M. Wang, W. W\'u, and D.-X. Yao, Bilayer two-orbital model of La$_3$Ni$_2$O$_7$ under pressure, Phys. Rev. Lett. \textbf{131}, 126001 (2023).
	\bibitem{Zhang2023b}Y. Zhang, L.-F. Lin, A. Moreo, and E. Dagotto, Electronic structure, orbital-selective behavior, and magnetic tendencies in the bilayer nickelate superconductor La$_3$Ni$_2$O$_7$ under pressure, arXiv:2306.03231.
	\bibitem{Christiansson2023}V. Christiansson, F. Petocchi, and P. Werner, Correlated electronic structure of La$_3$Ni$_2$O$_7$ under pressure, arXiv:2306.07931.
	\bibitem{Shilenko2023}D. A. Shilenko and I. V. Leonov, Correlated electronic structure, orbital-selective behavior, and magnetic correlations in double-layer La$_3$Ni$_2$O$_7$ under pressure, Phys. Rev. B \textbf{108}, 125105 (2023).
	\bibitem{Wu2023a}W. W\'u, Z. Luo, D.-X. Yao, and M. Wang, Charge transfer and Zhang-Rice singlet bands in the nickelate superconductor $\mathrm{La_3Ni_2O_7}$ under pressure, arXiv:2307.05662.
	\bibitem{Cao2023}Y. Cao and Y.-F. Yang, Flat bands promoted by Hund's rule coupling in the candidate double-layer high-temperature superconductor La$_3$Ni$_2$O$_7$, arXiv:2307.06806.
	\bibitem{Chen2023}X. Chen, P. Jiang, J. Li, Z. Zhong, and Y. Lu, Critical charge and spin instabilities in superconducting La$_3$Ni$_2$O$_7$, arXiv:2307.07154.
	
	
	\bibitem{Yang2023a}Q. G. Yang, H. Y. Liu, D. Wang, and Q. H. Wang, Possible $s^{\pm}$-wave superconductivity in La$_3$Ni$_2$O$_7$, arXiv: 2306.03706.
	\bibitem{Lechermann2023}F. Lechermann, J. Gondolf, S. B\"otzel, and I. M. Eremin, Electronic correlations and superconducting instability in La$_3$Ni$_2$O$_7$ under high pressure, arXiv:2306.05121.
	\bibitem{Sakakibara2023}H. Sakakibara, N. Kitamine, M. Ochi, and K. Kuroki, Possible high $T_c$ superconductivity in La$_3$Ni$_2$O$_7$ under high pressure through manifestation of a nearly-half-filled bilayer Hubbard model, arXiv:2306.06039.
	\bibitem{Gu2023}Y. Gu, C. Le, Z. Yang, X. Wu, and J. Hu, Effective model and pairing tendency in bilayer Ni-based superconductor La$_3$Ni$_2$O$_7$, arXiv:2306.07275.
	\bibitem{Shen2023}Y. Shen, M. Qin, and G.-M. Zhang, Effective bi-Layer model Hamiltonian and density-matrix renormalization group study for the high-$T_c$ superconductivity in La$_{3}$Ni$_{2}$O$_{7}$ under high pressure, arXiv:2306.07837.
	 \bibitem{Liu2023b}Y.-B. Liu, J.-W. Mei, F. Ye, W.-Q. Chen, and F. Yang, The $s^\pm$-Wave pairing and the destructive role of apical-oxygen deficiencies in La$_3$Ni$_2$O$_7$ under pressure, arXiv:2307.10144.
	\bibitem{Lu2023c}C. Lu, Z. Pan, F. Yang, and C. Wu, Interlayer coupling driven high-temperature superconductivity in La$_3$Ni$_2$O$_7$ under pressure, arXiv:2307.14965.
	\bibitem{Zhang2023d}Y. Zhang, L.-F. Lin, A. Moreo, T. A. Maier, and E. Dagotto, Structural phase transition, $s_{\pm}$-wave pairing and magnetic stripe order in the bilayered nickelate superconductor La$_3$Ni$_2$O$_7$ under pressure, arXiv:2307.15276.
	\bibitem{Oh2023}H. Oh and Y.-H. Zhang, Type II $t$-$J$ model and shared antiferromagnetic spin coupling from Hund’s Rule in superconducting La$_3$Ni$_2$O$_7$, arXiv:2307.15706.
	\bibitem{Liao2023}Z. Liao, L. Chen, G. Duan, Y. Wang, C. Liu, R. Yu, and Q. Si, Electron correlations and superconductivity in La$_3$Ni$_2$O$_7$ under pressure tuning, arXiv:2307.16697.
	\bibitem{Qu2023}X.-Z. Qu, D.-W. Qu, J. Chen, C. Wu, F. Yang, W. Li, and G. Su, Bilayer $t$-$J$-$J_\perp$ model and magnetically mediated pairing in the pressurized nickelate La$_3$Ni$_2$O$_7$, arXiv:2307.16873.
	\bibitem{Yang2023b}Y.-F. Yang, G.-M. Zhang, and F.-C. Zhang, Minimal effective model and possible high-$T_{c}$ mechanism for superconductivity of La$_{3}$Ni$_{2}$O$_{7}$ under high pressure, arXiv:2308.01176.
	\bibitem{Jiang2023a} K. Jiang, Z. Wang, and F.-C. Zhang, High temperature superconductivity in La$_3$Ni$_2$O$_7$, arXiv:2308.06771.
		\bibitem{Zhang2023e}Y. Zhang, L.-F. Lin, A. Moreo, T. A. Maier, and E. Dagotto, Trends of electronic structures and $s_{\pm}$-Wave pairing for the rare-earth series in bilayer nickelate superconductor $R_ 3$Ni$_2$O$_7$,arXiv:2308.07386.
	\bibitem{Huang2023}J. Huang, Z. D. Wang, and T. Zhou, Impurity and vortex states in the bilayer high-temperature superconductor $\mathrm{La}_3\mathrm{Ni}_2\mathrm{O}_7$, arXiv:2308.07651.

	
	\bibitem{Berg2008} E. Berg, D. Orgad, and S. A. Kivelson, Route to high-temperature superconductivity in composite systems, Phys. Rev. B  \textbf{78}, 094509 (2008).
	
	\bibitem{Mayr2005a} M. Mayr, G. Alvarez, C. Şen, and E. Dagotto, Phase fluctuations in strongly coupled $d$-wave superconductors, Phys. Rev. Lett. \textbf{94}, 217001 (2005).
	\bibitem{Dubi2007}Y. Dubi, Y. Meir, and Y. Avishai, Nature of the superconductor-insulator transition in disordered superconductors, Nature \textbf{449}, 876 (2007).
	\bibitem{Karmakar2020}M. Karmakar, Pauli limited $d$-wave superconductors: quantum breached pair phase and thermal transitions, J. Phys. Condens. Matter \textbf{32}, 405604 (2020).
	\bibitem{Pasrija2016}K. Pasrija, P. B. Chakraborty, and S. Kumar, Effective Hamiltonian based Monte Carlo for the BCS to BEC crossover in the attractive Hubbard model, Phys. Rev. B \textbf{94}, 165150 (2016).
	\bibitem{Dong2021a}J.-J. Dong, D. Huang, and Y.-F. Yang, Mutual information, quantum phase transition, and phase coherence in Kondo systems, Phys. Rev. B \textbf{104}, L081115 (2021).
	\bibitem{Mukherjee2014}A. Mukherjee, N. D. Patel, S. Dong, S. Johnston, A. Moreo, and E. Dagotto, Testing the Monte Carlo-mean field approximation in the one-band Hubbard model, Phys. Rev. B  \textbf{90}, 205113 (2014).
	\bibitem{Liang2013}S. Liang, A. Moreo, and E. Dagotto, Nematic state of pnictides stabilized by interplay between spin, orbital, and lattice degrees of freedom, Phys. Rev. Lett. \textbf{111}, 047004 (2013).
	
	\bibitem {Coleman2015} P. Coleman, \textit{Introduction to Many-body Physics}, (Cambridge University Press, Cambridge, U.K., 2015).
	
	
	\bibitem{Qin2023PRB} Q. Qin, J.-J. Dong, Y. Sheng, D. Huang, and Y.-F. Yang, Superconducting fluctuations and charge-4$e$ plaquette state at strong coupling, Phys. Rev. B \textbf{108}, 054506 (2023).
	\bibitem{Han2010}Q. Han, T. Li, and Z. D. Wang, Pseudogap and Fermi-arc evolution in the phase-fluctuation scenario, Phys. Rev. B \textbf{82}, 052503 (2010).
	\bibitem{Zhong2011}Y. W. Zhong, T. Li, and Q. Han, Monte Carlo Study of thermal fluctuations and Fermi-arc formation in $d$-wave superconductors, Phys. Rev. B \textbf{84}, 024522 (2011).
	\bibitem{Singh2021}D. K. Singh, S. Kadge, Y. Bang, and P. Majumdar, Fermi arcs and pseudogap phase in a minimal microscopic model of $d$-wave superconductivity, Phys. Rev. B \textbf{105}, 054501 (2022).
	
	
	\bibitem{Dong2022PRB}J.-J. Dong and Y.-F. Yang, Development of long-range phase coherence on the Kondo lattice, Phys. Rev. B 106, L161114 (2022).
	
	\bibitem{Cover2006}T. M. Cover and J. A. Thomas, \textit{Elements of Information Theory}, Wiley Series in Telecommunications and Signal Processing (Wiley-Interscience, Hoboken, NJ, 2006).
	\bibitem{Kraskov2004PRE} A. Kraskov, H. St\"{o}gbauer, and P. Grassberger, Estimating mutual information, Phys. Rev. E \textbf{69}, 066138 (2004).
	\bibitem{Varanasi1999}	G. A. Darbellay and I. Vajda, Estimation of the Information by an Adaptive Partitioning of the Observation Space, IEEE Trans. Inf. Theory \textbf{45}, 1315 (1999).
	\bibitem{Khan2007}S. Khan, S. Bandyopadhyay, A. R. Ganguly, S. Saigal, D. J. Erickson, V. Protopopescu, and G. Ostrouchov, Relative performance of mutual information estimation methods for quantifying the dependence among short and noisy data, Phys. Rev. E \textbf{76}, 026209 (2007).
	\bibitem {Belghazi2018PMLR} M. I. Belghazi, A. Baratin, S. Rajeswar, S. Ozair, Y. Bengio, A. Courville, and R. D. Hjelm, Mutual information neural estimation,
	in \textit{Proceedings of the 35th International Conference on Machine Learning, Stockholmsmässan, Stockholm Sweden,} edited by J. Dy and A. Krause 
	(PMLR, Stockholmsmässan, Stockholm Sweden, 2018), p. 531.
	\bibitem{Poolel2019PMLR} B. Poole, S. Ozair, A. V. D. Oord, A. A. Alemi, and G. Tucker, On variational bounds of mutual information, in \textit{Proceedings of the 36th International Conference on Machine Learning, Long Beach, California, USA,}  edited by K. Chaudhuri and R. Salakhutdinov 
	( PMLR, Long Beach, California, USA, 2019), p. 5171.
	
   \bibitem{Drouin-Touchette2022} V. Drouin-Touchette, The Kosterlitz-Thouless phase transition: an introduction for the intrepid student, arXiv:2207.13748.
	
	\bibitem{Berezinskii1972}V. L. Berezinskii, Destruction of long-range order in one-dimensional and two-dimensional systems possessing a continuous symmetry group. II. Quantum Systems,Zh. Eksp. Teor.
	Fiz. \textbf{61}, 1144 (1971) [Sov. Phys. JETP \textbf{34}, 610 (1972)].
	\bibitem{Kosterlitz1973}J. M. Kosterlitz and D. J. Thouless, Ordering, metastability and phase transitions in two-dimensional systems, J. Phys. C: Solid State Phys. \textbf{6}, 1181 (1973).
	\bibitem{Kosterlitz1974}J. M. Kosterlitz, The critical properties of the two-dimensional XY model, J. Phys. C: Solid State Phys. \textbf{7}, 1046 (1974).
	
	
	\bibitem{Keimer2015} B. Keimer, S. A. Kivelson, M. R. Norman, S. Uchida, and J. Zaanen, From quantum matter to high-temperature superconductivity in copper oxides, Nature \textbf{518}, 179 (2015).
	\bibitem{Emery1995}V. J. Emery and S. A. Kivelson, Importance of phase fluctuations in superconductors with small superfluid density, Nature \textbf{374}, 434 (1995).
	
	\bibitem{Yoshida2009} T. Yoshida, M. Hashimoto, S. Ideta, A. Fujimori, K. Tanaka, N. Mannella, Z. Hussain, Z. X. Shen, M. Kubota, K. Ono, S. Komiya, Y. Ando, H. Eisaki, and S. Uchida, Universal versus material-dependent two-gap behaviors of the high-$T_c$ cuprate superconductors: angle-resolved photoemission study of La$_{2-x}$Sr$_x$CuO$_4$, Phys. Rev. Lett. \textbf{103}, 037004 (2009).
	\bibitem{Richter2013} C. Richter, H. Boschker, W. Dietsche, E. Fillis-Tsirakis, R. Jany, F. Loder, L. F. Kourkoutis, D. A. Muller, J. R. Kirtley, C. W. Schneider, and J. Mannhart, Interface superconductor with gap behaviour like a high-temperature superconductor, Nature \textbf{502}, 528 (2013).
	
	\bibitem{Kim2022}H. Kim, Y. Choi, C. Lewandowski, A. Thomson, Y. Zhang, R. Polski, K. Watanabe, T. Taniguchi, J. Alicea, and S. Nadj-Perge, Evidence for unconventional superconductivity in twisted trilayer graphene, Nature \textbf{606}, 494 (2022).
   \bibitem{Yao2019}G. Yao, M. C. Duan, N. Liu, Y. Wu, D. D. Guan, S. Wang, H. Zheng, Y. Y. Li, C. Liu, and J. F. Jia, Diamagnetic response of potassium-adsorbed multilayer FeSe film, Phys. Rev. Lett. \textbf{123}, 257001 (2019).
	\bibitem{Inosov2011} D. S. Inosov, J. T. Park, A. Charnukha, Y. Li, A. V. Boris, B. Keimer, and V. Hinkov, Crossover from weak to strong pairing in unconventional superconductors, Phys. Rev. B \textbf{83}, 214520 (2011). 

	
	
\end{thebibliography}
\end{document}